\documentclass[a4paper,11pt]{article}
\usepackage{pos}
\usepackage[utf8]{inputenc}
\usepackage[T1]{fontenc}
\usepackage[british]{babel}
\usepackage{%
    graphicx,
    subcaption,
    multirow,
    booktabs,
    xifthen,
    pdflscape,
    longtable
}

\usepackage{cleveref}

\graphicspath{{figures/}}
\newsavebox{\largestimage} % Used to vertically center subfigures
\crefname{section}{section}{sections}
\Crefname{section}{Section}{Sections}
\crefname{figure}{figure}{figures}
\Crefname{figure}{Figure}{Figures}
\crefname{table}{table}{tables}
\Crefname{table}{Table}{Tables}
\crefname{appendix}{appendix}{appendices}
\Crefname{appendix}{Appendix}{Appendices}
\crefname{equation}{equation}{equations}
\Crefname{equation}{Eq.}{Eqs.}
\crefname{enumi}{step}{steps}
\Crefname{enumi}{Step}{Steps}

\newcommand{\coeff}[2]{\mathcal{\uppercase{#1}}_{#2}}

\newcommand{\crit}{c}
\newcommand{\tric}{\text{tric}}

\newcommand{\clqcd}{\texttt{CL\kern-.25em\textsuperscript{2}QCD}}
\newcommand{\Ocl}{OpenCL}
\newcommand{\Amd}{AMD}
\newcommand{\bahamas}{\texttt{BaHaMAS}}

\newcommand{\NSigma}{N_\sigma}
\newcommand{\NTau}[1][]{%
    \ifthenelse{\isempty{#1}}%
      {N_{\tau}}%
      {N_{\tau_{#1}}}%
}
\newcommand{\Nf}{N_\text{f}}
\newcommand{\T}{T}
\newcommand{\mud}{m_{u,d}}

\newcommand{\Tc}{\T_{\crit}}
\newcommand{\mc}{m_{\crit}}

\newcommand{\Action}{\mathcal S}

\title{The QCD chiral phase transition for different numbers of quark flavours}
\ShortTitle{QCD chiral phase transition}

\author[a]{Francesca Cuteri}
\author*[a,b]{Owe Philipsen}
\author[a,c]{Alessandro Sciarra}

\affiliation[a]{ITP, Goethe Universit\"at Frankfurt,\\
  Max-von-Laue-Str. 1, 60438 Frankfurt, Germany}
  
\affiliation[b]{John von Neumann Institute for Computing (NIC)\\
at GSI, Planckstr.\ 1, 64291 Darmstadt, Germany}

\affiliation[c]{Frankfurt Institute for Advanced Studies (FIAS)-Goethe University,\\
Ruth-Moufang-Str. 1, 60438 Frankfurt am Main, Germany}

\emailAdd{cuteri@itp.uni-frankfurt.de}
\emailAdd{philipsen@itp.uni-frankfurt.de}
\emailAdd{sciarra@itp.uni-frankfurt.de}

\abstract{
We present results from a comprehensive study of the location of the chiral critical surface, which separates regions of 
first-order chiral transitions from analytic crossovers, in the bare parameter space of lattice QCD with unimproved
staggered fermions. We study the theories with $\Nf\in[2,8]$ and trace the 
chiral critical surface along diminishing lattice spacing, with $\NTau=\{4,6,8\}$. This allows for an extrapolation to the lattice chiral limit,
where the surface has to terminate in a tricritical line, employing known tricritical scaling relations. Knowing the phase structure in 
the lattice bare parameter space allows to draw conclusions for the approach to the continuum and chiral limits taken in 
the appropriate order. Our data provide evidence for the continuum chiral limit to feature a second-order transition for all 
$\Nf\in[2,7]$. We perform an analogous scaling analysis with already published data from $\Nf=3$ $O(a)$-improved Wilson 
fermions, which is also consistent with a second-order transition in the continuum. A modified Columbia plot
reflecting those results is suggested. 
}

\FullConference{%
 The 38th International Symposium on Lattice Field Theory, LATTICE2021
  26th-30th July, 2021
  Zoom/Gather@Massachusetts Institute of Technology
}

\begin{document}
\maketitle

\section{Introduction}

\begin{figure}[t]
    \centering
    \subcaptionbox{First-order scenario for $\Nf=2, \mud=0$. \label{fig:columbia-first}}{\includegraphics[width=0.48\textwidth]{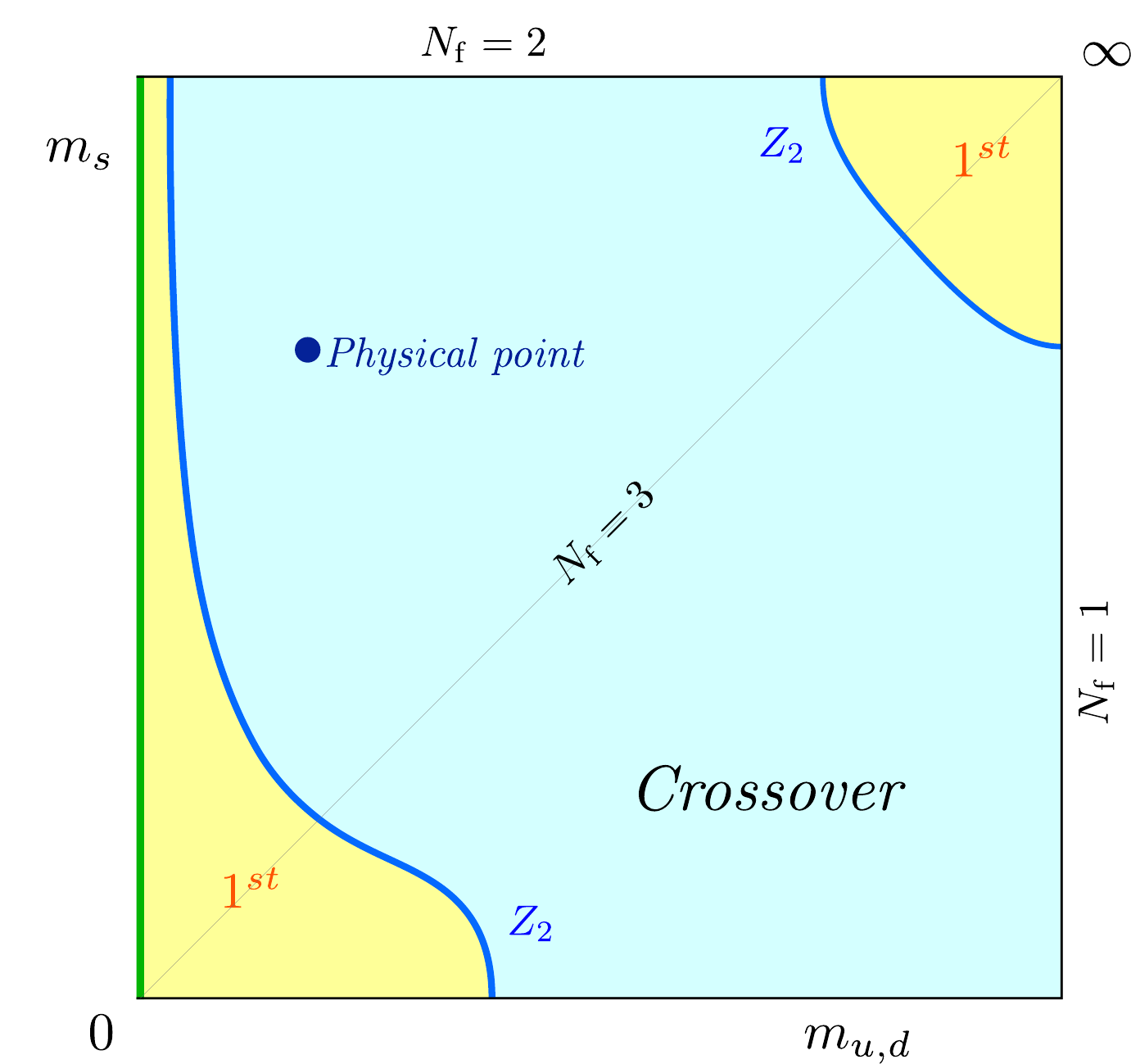}}
    \hfill
    \subcaptionbox{Second-order scenario for $\Nf=2, \mud=0$.\label{fig:columbia-second}}{\includegraphics[width=0.48\textwidth]{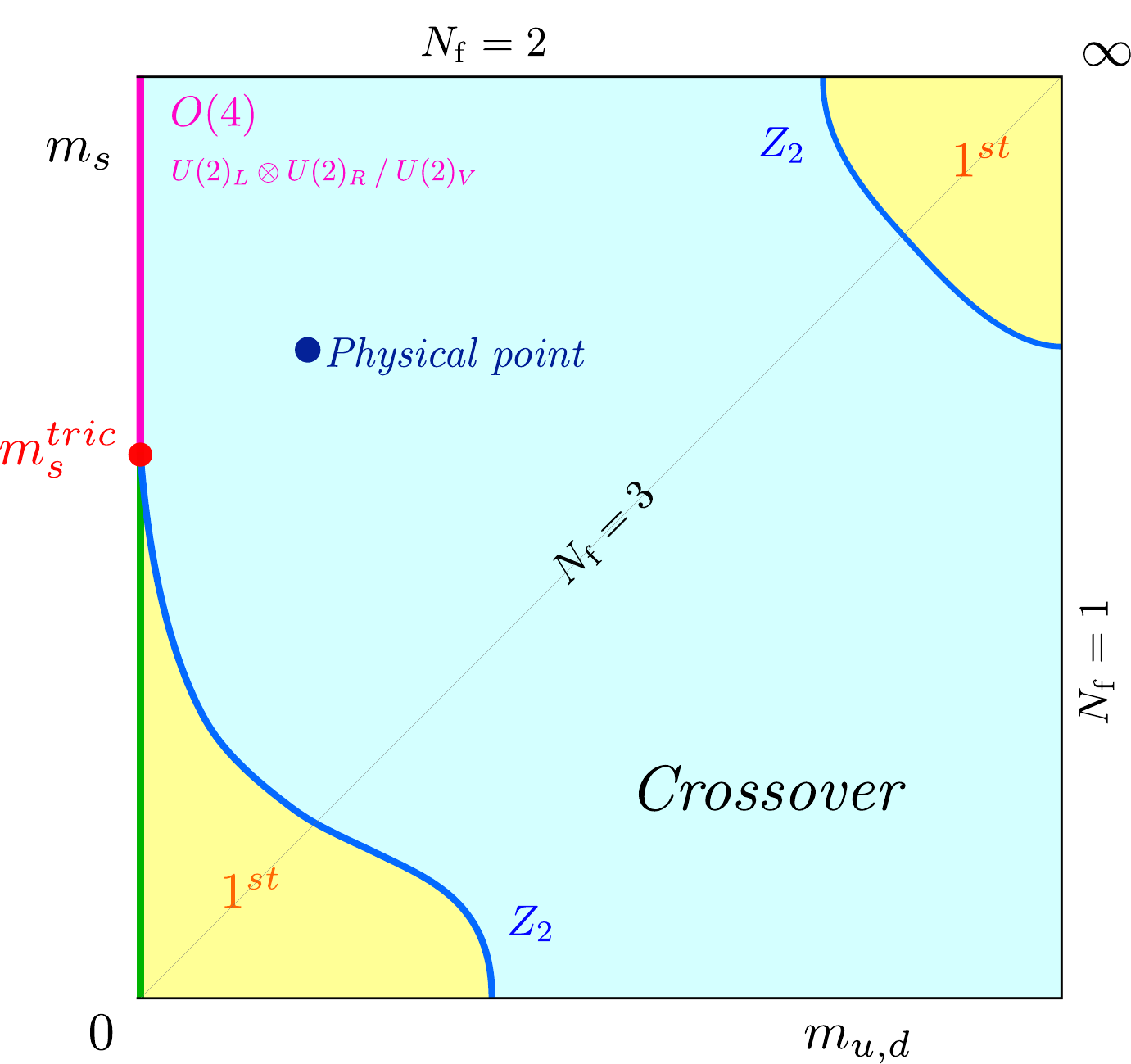}}
    \caption{%
      Possible scenarios for the order of the thermal QCD transition as a function of the quark masses.
         }
    \label{fig:columbia}
\end{figure}
The nature of the QCD chiral phase transition in the limit of massless quarks has been a longstanding, open problem. Its 
unambiguous, non-perturbative resolution is important, because the light quarks
in nature are close to the chiral limit. This raises the question %whether QCD thermodynamics is also affected
%by the chiral limit, and in particular 
whether traces of the chiral phase transition might be detectable experimentally.
Unfortunately, direct lattice simulations of the chiral limit are not feasible due to the singular nature of the fermion determinant,
so that extrapolations are inevitable, introducing systematic errors.

The nature of the thermal QCD transition with $\Nf=2+1$ flavours is usually displayed as a function of the quark masses in 
a Columbia plot \cite{Brown:1990ev}, as in \cref{fig:columbia}. The two possibilities correspond to the predictions of
the renormalisation group flow in 3D sigma models, augmented by a 't Hooft term for the axial anomaly, using the epsilon 
expansion \cite{Pisarski:1983ms}: for $\Nf\geq 3$ the chiral phase transition in the massless limit is predicted to be of first order, whereas 
for $\Nf=2$ it depends on whether the axial anomaly is effectively restored at the critical temperature (first order), or remains
broken (second order). Numerous numerical lattice investigations have been devoted to determine the location of
the second-order boundaries to distinguish between these scenarios. 
One generally observes widely differing values for the pseudo-scalar mass evaluated on the critical boundary at different points
and between different actions. Unimproved staggered and Wilson actions as well as $O(a)$-improved Wilson actions 
on coarse lattices see a first-order region both for $\Nf\in\{2,3\}$, but it is found to shrink drastically as the lattice is made finer
\cite{Iwasaki:1996zt,Karsch:2001nf,deForcrand:2003vyj,deForcrand:2007rq,Jin:2014hea,Bonati:2014kpa,Philipsen:2016hkv,Cuteri:2017gci,Jin:2017jjp,Kuramashi:2020meg}.
On the other hand, improved staggered actions do not see a first-order region at all 
down to $m_{PS}\approx 50$ MeV \cite{Bazavov:2017xul}. 
For a detailed review, see \cite{Philipsen:2021qji}. The question is whether there are actual contradictions between 
different discretisations, or whether they all converge towards one answer, and whether the correct answer has
any first-order region in the continuum limit.
Here we discuss a novel way to analyse the cutoff effects associated with the observed first-order region, which 
suggests a unified description of all available lattice results.

\section{The Columbia plot for mass-degenerate quarks and tricritical scaling}

\begin{figure}[t]
    \centering
    \subcaptionbox{First-order scenario for $\Nf=2, \mud=0$. \label{fig:columbia-first}}{\includegraphics[width=0.48\textwidth]{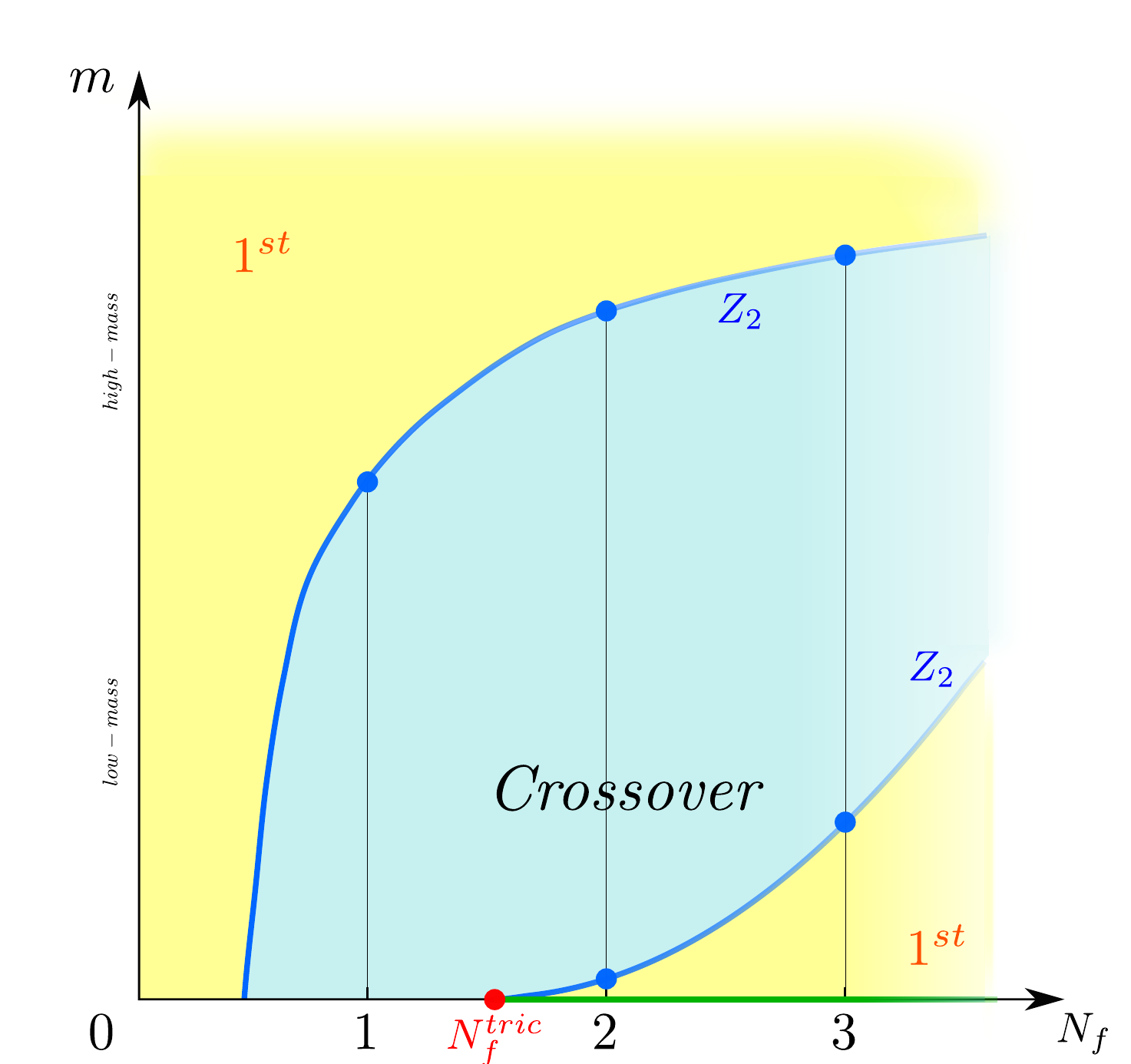}}
    \hfill
    \subcaptionbox{Second-order scenario for $\Nf=2, \mud=0$.\label{fig:columbia-second}}{\includegraphics[width=0.48\textwidth]{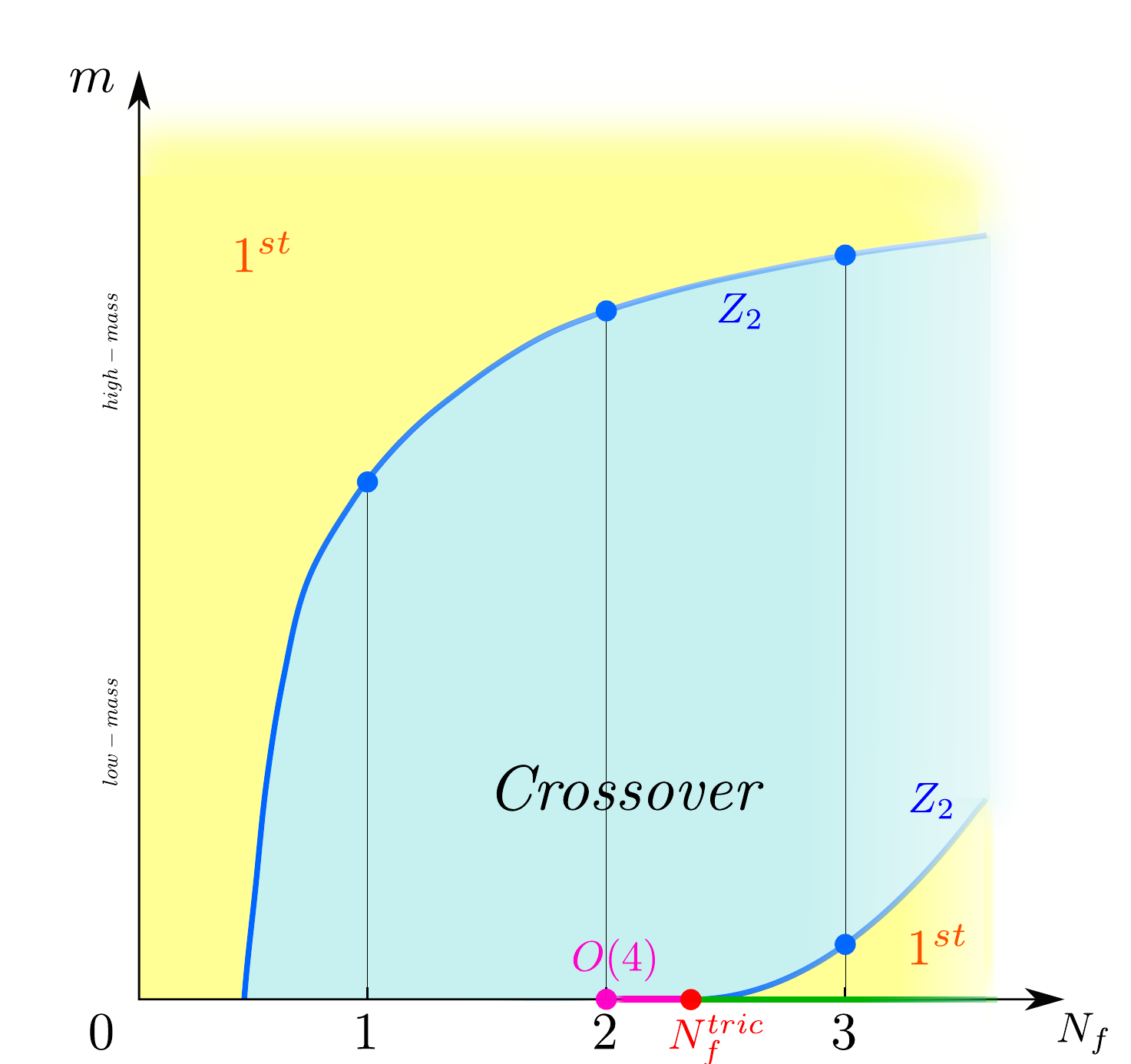}}
    \caption{%
      Scenarios for the order of the thermal QCD transition as a function of $\Nf$ degenerate flavours \cite{Cuteri:2017gci}.
         }
    \label{fig:columbia2}
\end{figure}
The analysis of the chiral phase transition is facilitated by
considering only mass-degenerate quarks with the partition function (in continuum notation)
\begin{equation}
 Z(\Nf,g,m)=\int {\cal D}A_\mu \; (\det M[A_\mu,m])^{\Nf}\; e^{-\Action_\text{YM}[A_\mu]}\;.
\end{equation}
This can formally be viewed as a statistical system depending on a continuous parameter $\Nf$, which allows for an
alternative interpolation between $\Nf=2$ and $\Nf=3$, rather than varying the strange quark mass.
% Moreover, this readily generalises to any number of flavours. 
The Columbia plots  of \cref{fig:columbia} then translate to the analogous versions of 
\cref{fig:columbia2}. Looking at the problem from this perspective offers two important benefits: First, since there is no chiral phase
transition for $\Nf=1$, a first-order transition in the chiral limit for any $\Nf>1$ must necessarily weaken with decreasing $\Nf$,
until it vanishes in a tricritical point. This is  because a first-order transition in the chiral limit represents a coexistence of three states,
with $\pm\langle\bar{\psi}\psi\rangle\neq 0$ and $\langle \bar{\psi}\psi\rangle=0$,
and the point where the diminishing latent heat vanishes is tricritical. Hence, both the second-order \textit{and} the first-order scenario
for $\Nf=2$ now feature a tricritical point in the Columbia plot, with either $2<\Nf^\mathrm{tric}<3$ or $1<\Nf^\mathrm{tric}<2$, 
respectively. Second, the $Z_2$-critical line, which separates the parameter region with analytic crossover from that of first-order transitions,
enters the tricritical point as a function of the symmetry breaking scaling field $(am)^{2/5}$,
\begin{equation}
\Nf^c(am)=\Nf^\mathrm{tric} + \coeff{A}{1} \;(am)^{2/5} + \coeff{A}{2}\;(am)^{4/5}+ \ldots \;.
\label{eq:scale}
\end{equation}
The critical exponents of the scaling field take known mean field values, since the upper critical dimension for a tricritical point is 
three \cite{lawrie}. On the lattice, there will be an additional dependence on $\NTau$, viz.~the lattice spacing, and hence a tricritical line. Based on these facts, and with explicit first-order transitions seen on coarse lattices, 
our task is now reduced to locating the chiral intercept $\Nf^\mathrm{tric}(\NTau)$ based on a polynomial with known exponents,
rather than having to distinguish between different sets of critical exponents.

\section{Simulations and analysis}

For our numerical investigation, we use the standard unimproved Wilson gauge and staggered fermion actions.
All numerical simulations have been performed using the publicly available \Ocl-based code \clqcd, which is optimised to run efficiently on \Amd{} GPUs and contains an implementation of the RHMC algorithm for unimproved rooted staggered fermions.
Version \texttt{v1.0}~\cite{pinke_cl2qcd_2018} has been employed for simulations on smaller $\NTau$ on the L-CSC supercomputer at GSI, while version \texttt{v1.1}~\cite{sciarra_cl2qcd_2021} has been run on the HLR supercomputer at Goethe University 
to run the most costly simulations.
The thousands of necessary simulations were efficiently handled by the \bahamas\ software~\cite{sciarra_bahamas_2021}.

Our method to determine the order of the chiral transition by finite size scaling is standard. 
We evaluate the chiral condensate $\langle \bar{\psi}\psi\rangle$,
which becomes an exact order parameter in the massless limit, and its standardised cumulants $B_{3,4}$ defined as
\begin{equation}
    B_n(\beta,am, \Nf, \NTau, \NSigma) =
    \frac{\left\langle\left(\bar{\psi}\psi - \left\langle\bar{\psi}\psi\right\rangle\right)^n\right\rangle}{\left\langle\left(\bar{\psi}\psi - \left\langle\bar{\psi}\psi\right\rangle\right)^2\right\rangle^{{n/2}^{\vphantom{x}}}} \;.
\end{equation}
For any fixed volume, the bare parameter space of unimproved staggered fermions is four-dimensional, $(\beta,am,\Nf,\NTau)$. 
We first locate the
phase boundary between chirally broken and restored regions by the condition of vanishing skewness for the
distribution of the chiral condensate, typically by scanning in $\beta$, $B_3(\beta_c,am,\Nf,\NTau,\NSigma)=0$. 
This defines a three-dimensional subspace, which is composed of a region of crossover transtions
and a region of first-order transitions. These are separated by a $Z_2$-critical surface, to be identified by the parameter values
where the kurtosis assumes its 3D Ising value, 
$B_4(\beta_c,am_c,\Nf,\NTau,\NSigma=\infty)=1.604$. 
On finite but sufficiently large volumes close to the thermodynamic limit, 
the kurtosis can be expanded about the critical point,
\begin{equation}
B_4(\beta_c,am,\Nf,\NTau,\NSigma)=1.604+ \coeff{B}{1}(\beta_c,\Nf,\NTau)\; (am-am_c)\NSigma^{1/\nu}+\ldots \;,
\end{equation} 
through which it passes smoothly. As the volume is increased, this approaches a step function and the rate of
the approach to the thermodynamic limit is governed by a 3D Ising critical exponent, $\nu=0.6301$. Dots in the 
equation indicate additional terms that vanish in the infinite volume limit. For the set of aspect ratios $\NSigma/\NTau\in\{2,3,4,5\}$ used throughout,
the corrections were found to be statistically insignificant in most cases, so that fits to this equation provide estimates
for the critical masses, and hence the location of the $Z_2$-critical surface in infinite volume. 

For each parameter combination, we generally simulated four independent Monte Carlo chains 
until their $B_4$-values agreed to within three standard deviations or better, upon which they were merged. In order to tune
precisely to the phase boundary, the multi-histogram method was used to interpolate between simulated $\beta$-values \cite{FSReweighting}. 
All steps of our analysis follow the details described in \cite{Cuteri:2017gci,Cuteri:2020yke}. In addition to those, we 
have implemented a new error analysis for the critical masses $am_c$ based on bootstrap estimators, which is entirely
independent of the usual $\chi^2$-minimisation. Besides providing a crucial reliability check on the fits, 
this procedure typically produces slightly smaller errors \cite{Cuteri:2021ikv}. Altogether, the following results are based on 
120 million Monte Carlo trajectories spread over 600 different parameter combinations, obtained over a span of several years.

\section{The bare parameter space of staggered lattice QCD}

\begin{figure}[t]
    \centering
    \subcaptionbox{Critical mass in units of $T$. Lines represent next-to-leading order scaling fits to \cref{eq:mscale} \cite{Cuteri:2021ikv}. \label{fig:m-nf_all}}{\includegraphics[width=0.48\textwidth]{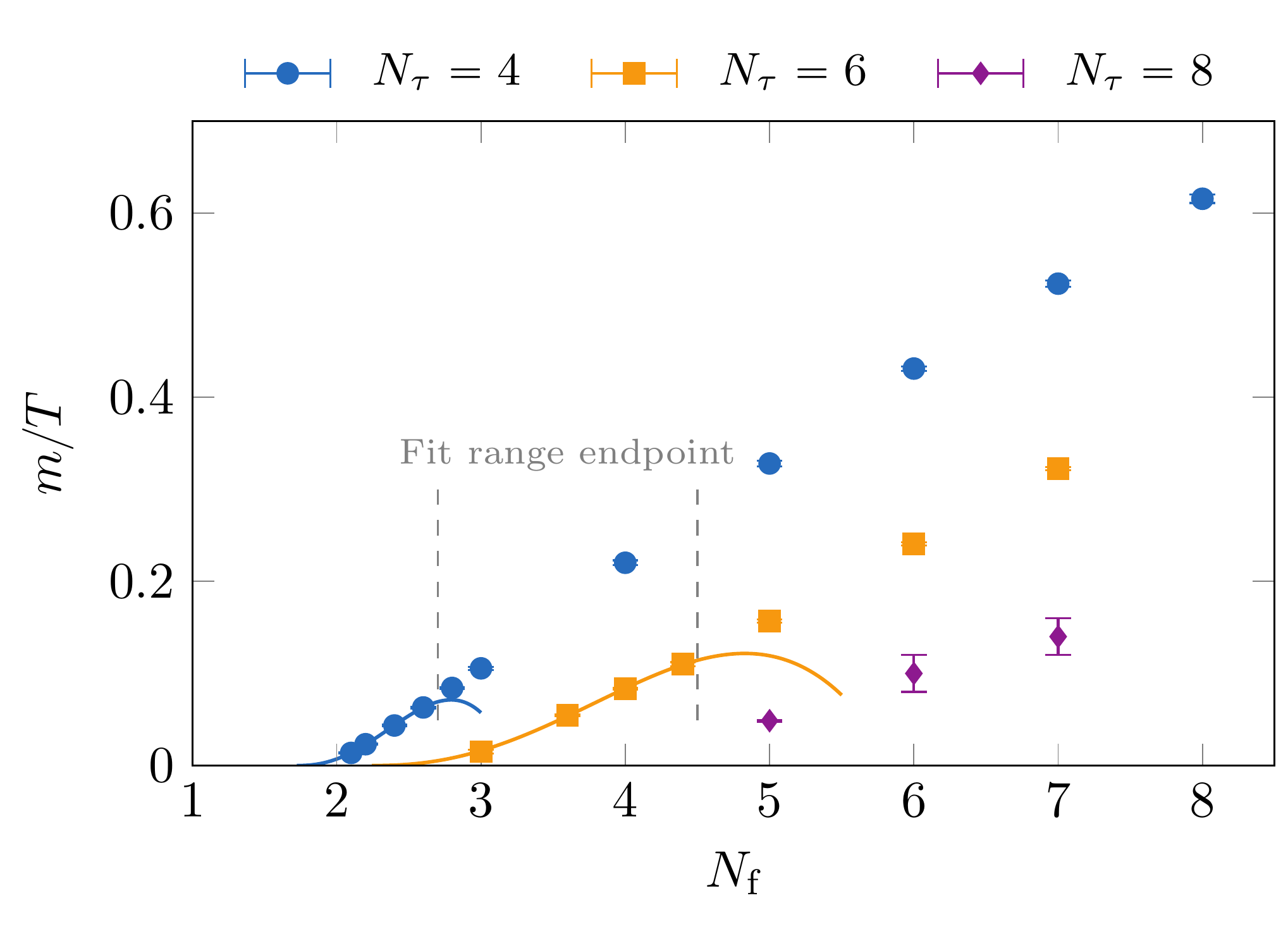}}
    \hfill
    \subcaptionbox{Critical mass in lattice units, with a leading-order scaling fit to \cref{eq:mscale}. 
    From \cite{Cuteri:2017gci}. \label{fig:m-nf_4}}{\includegraphics[width=0.48\textwidth]{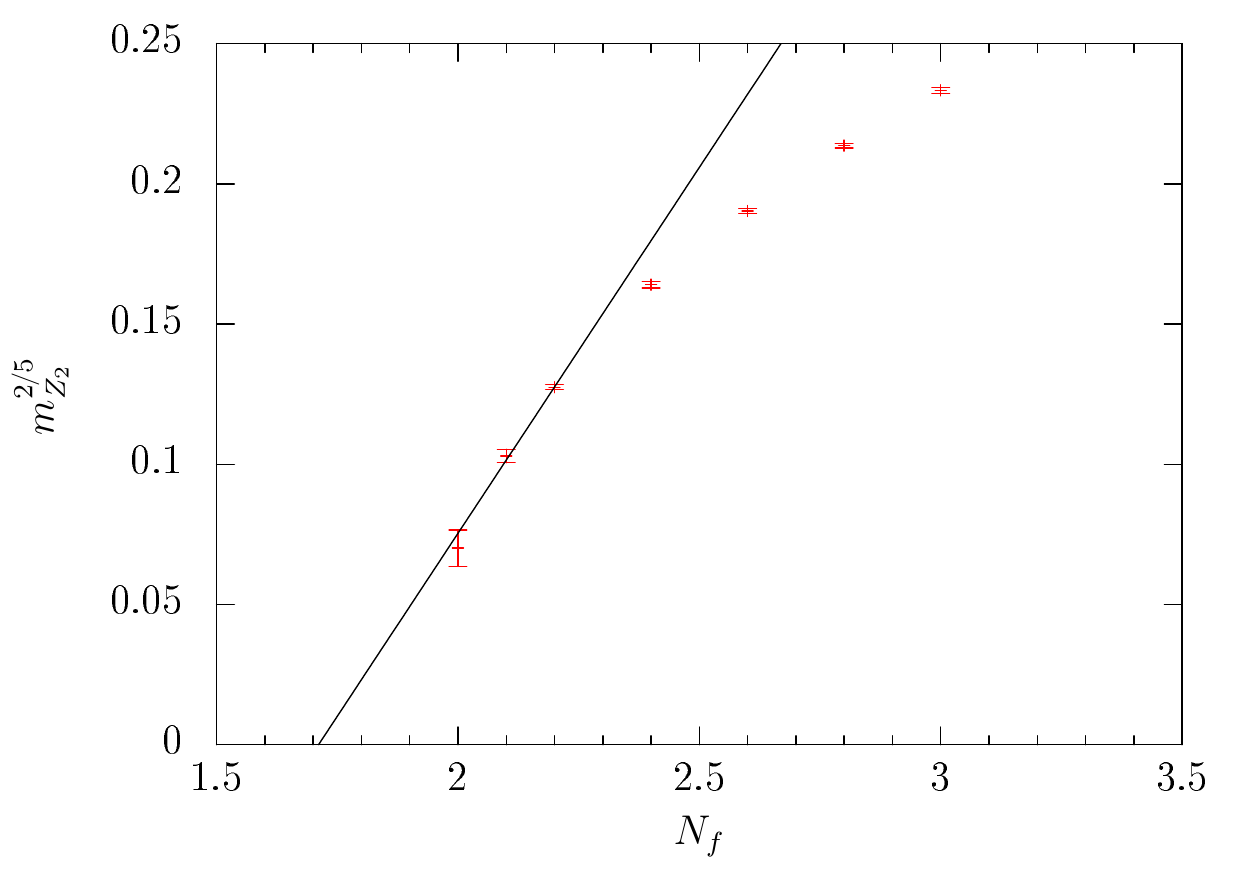}}
    \caption{% 
      The chiral critical surface projected onto the $(am,\Nf)$-plane. Regions above the lines represent crossover transitions, those
      below first-order transitions.
         }
    \label{fig:m-nf}
\end{figure}
The result of our finite size scaling analysis is the location of the chiral critical surface in the infinite volume bare parameter space
of the lattice theory. To analyse its implications, we study its projections onto all possible planes of variable pairings. 
We start with \cref{fig:m-nf_all}, which represents the lattice analogue for the sketched Columbia plots in \cref{fig:columbia2}.
The data represent the chiral critical surface corresponding to different fixed $\NTau$, separating crossover transitions above from
first-order transitions below.  One clearly observes a strenghtening of the first-order transition with increasing $\Nf$ and a weakening 
with increasing $\NTau$. Moreover, there is no sign of convergence towards a continuum limit yet. Thus, large parts of the
first-order region must be a cutoff effect, which is evidently stronger for larger $\Nf$. Our main interest is in the intercept of the
curves with the lattice chiral limit, i.e.~the tricritical line $\Nf^\mathrm{tric}(\NTau)$.
Tricritical scaling can be appreciated
in the rescaled \cref{fig:m-nf_4}, where earlier $\NTau=4$ data approach a leading order scaling 
relation~\cite{Cuteri:2017gci}.
This allows for an extrapolation
\begin{align}\label{eq:mscale}
  am_c\big(\Nf(\NTau),\NTau\big)
    &= \coeff{D}{1}(\NTau)\big( \Nf-\Nf^\mathrm{tric}(\NTau)\big)^{5/2}+
    \coeff{D}{2}(\NTau)\big( \Nf-\Nf^\mathrm{tric}(\NTau)\big)^{7/2}+ \ldots \;,
\end{align}
where we inverted \cref{eq:scale}, since the $\Nf$-values are exact while $am_c$ has errors. 
Note also that the $\Nf=2$ data point
 in  \cref{fig:m-nf_4} has been obtained by a tricritical extrapolation
 in imaginary chemical potential at fixed $\Nf=2$ \cite{Bonati:2014kpa}. This is an independent confirmation of the 
 bare quark mass as a tricritical scaling field near its chiral limit.
Unfortunately, the scaling region is small in this variable pairing. Next-to-leading order fits in the left figure
 predict $\Nf^\mathrm{tric}(\NTau=4)\approx 1.71(3)$ and 
 $\Nf^\mathrm{tric}(\NTau=6)\approx 2.20(8)$.  One concludes that, for~unimproved staggered fermions, 
 the $\Nf=2$ massless theory shows a first-order transition on $\NTau=4$, but~a second-order transition on all
 finer lattices and in the continuum. The question is what happens to the $\Nf\geq 3$ theories, since the $\NTau=8$ data
 suggest a further slide of the critical line towards larger $\Nf$.

\begin{figure}[t]
    \centering
    \begin{minipage}{0.46\textwidth}
       \centering
     %  \vspace*{-0.2cm}
       \includegraphics[width=\linewidth]{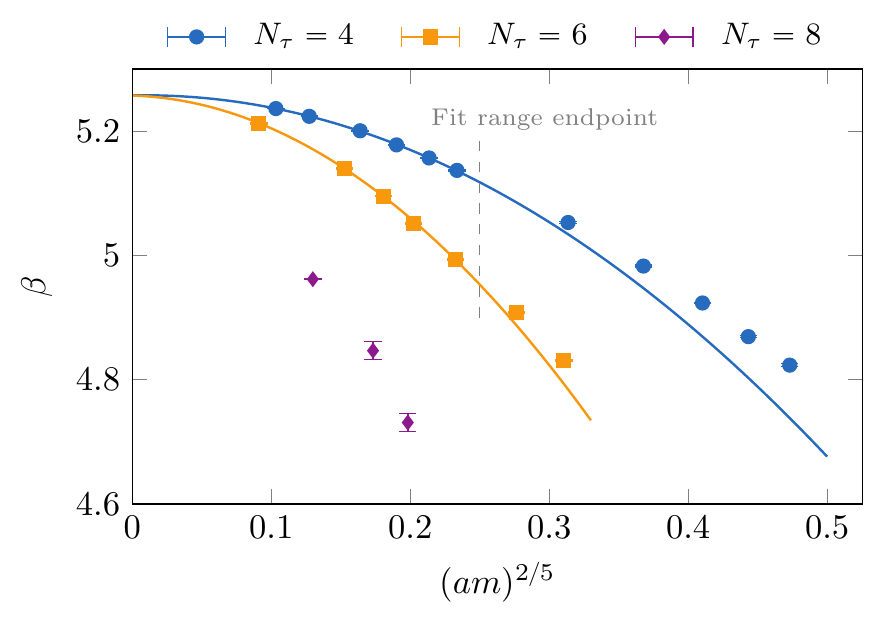}
       \captionof{figure}{
                Chiral critical surface projected onto the $(\beta,am)$-plane, fitted to next-to-leading order tricritical scaling. 
                From \cite{Cuteri:2021ikv}. 
            }\label{fig:beta-m}  
     \end{minipage}%
     \hfill
     \begin{minipage}{0.48\textwidth}
       \centering
       \vspace*{-2mm}
       \includegraphics[width=\linewidth]{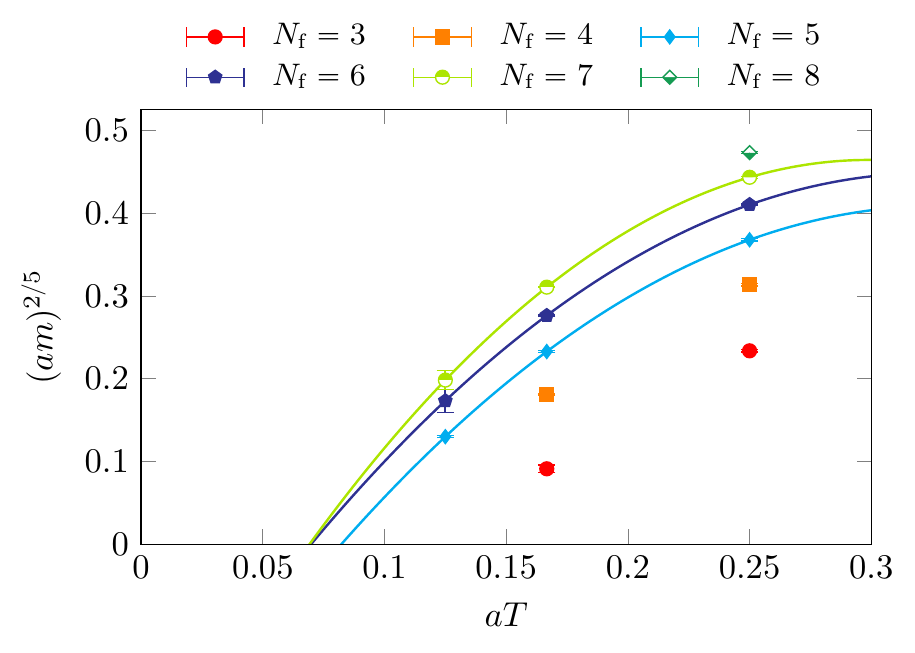}
       \captionof{figure}{
            Chiral critical surface projected onto the $(am,aT=\NTau^{-1}))$-plane, interpolated by next-to-leading order tricritical scaling.
                From \cite{Cuteri:2021ikv}.    
            }\label{fig:m-nt}  
     \end{minipage}
\end{figure}

More information can be obtained by analysing the same data in different variable pairings. 
First, \cref{fig:beta-m} clearly confirms the tricritical scaling behaviour, which does not get superseded
by the linear $\Nf$-dependence observed in \cref{fig:m-nf_all}. The critical lines are implicitly parametrised by $\Nf$, 
the region above them corresponds to
crossover transitions, and those below to first-order transitions. The lower $\beta$-axis represents a first-order triple line
that ends in the tricritical point marked by the intercept of the curves with the lattice chiral limit. 

\begin{figure}[t]
    \centering
    \subcaptionbox{Fits to \cref{eq:m_T_1st}, corresponding to the first-order scenario. From \cite{Cuteri:2021ikv}. \label{fig:nf5-1st}}{\includegraphics[width=0.48\textwidth]{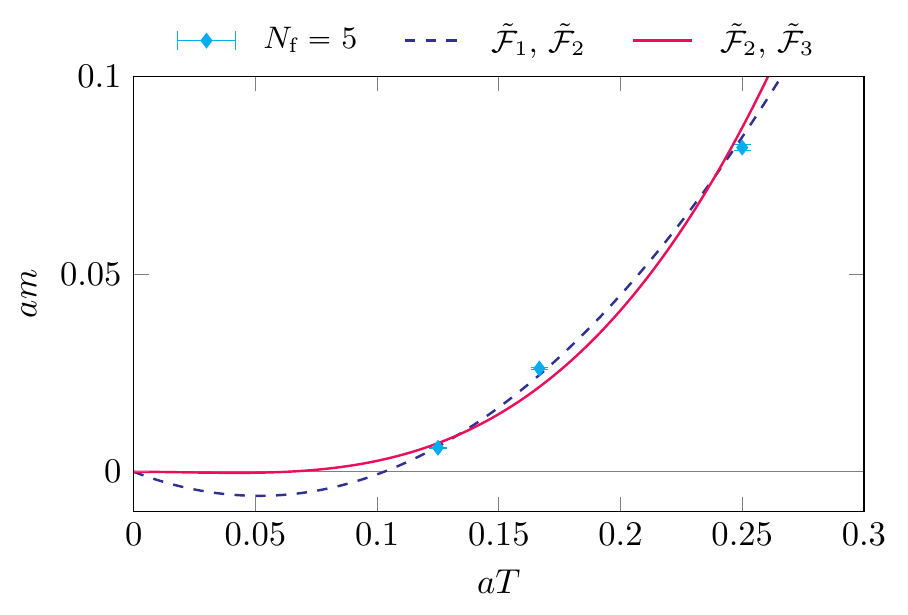}}
    \hfill
    \subcaptionbox{Fits to \cref{eq:T_m,eq:m_T}, corresponding to the second-order scenario. From \cite{Cuteri:2021ikv}. \label{fig:nf5-2nd}}{\includegraphics[width=0.48\textwidth]{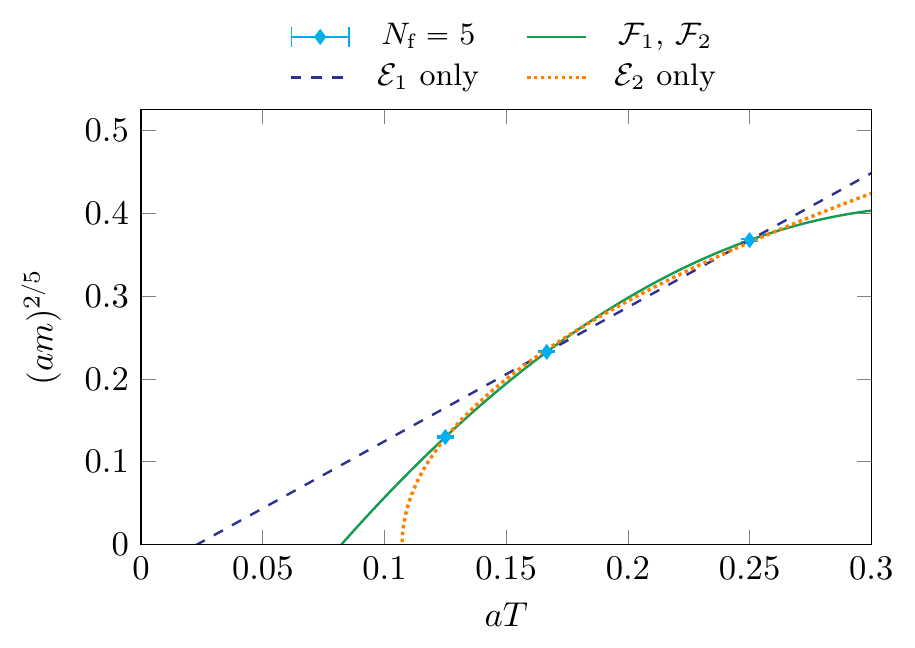}}
    \caption{%
      Fitting and extrapolating the chiral critical line of the $\Nf=5$ theory.
         }
    \label{fig:nf5}
\end{figure}

In \cref{fig:m-nt} we show the rescaled
critical bare quark masses plotted as a function of $\NTau$.  
Only a slight curvature is exhibited by those $\Nf$ with three data points, which thus are compatible
with next-to-leading order scaling and a tricritical point at some finite $\NTau^\mathrm{tric}(\Nf)$. 
Note however, that for fixed $\Nf$-values a tricritical
point is not guaranteed to exist, and one must test for the functional behaviour. As an example, \cref{fig:nf5} shows fits to
the $\Nf=5$ data assuming different scenarios. For a first-order chiral transition there is 
a finite continuum critical mass $m_c$, modified by the usual polynomial discretisation effects, so that in lattice units one has
\begin{equation}\label{eq:m_T_1st}
    a\mc(\NTau,\Nf)=\coeff{\tilde{F}}{1}(\Nf)\; aT+ \coeff{\tilde{F}}{2}(\Nf)\;(aT)^2 + \coeff{\tilde{F}}{3}(\Nf)\;(aT)^3 + \ldots \;.
\end{equation}
Two different next-to-leading order fits shown in \cref{fig:nf5-1st} have $\chi^2_\mathrm{dof}>50$ and visibly fail to describe the data.
 By contrast, a description with a tricritical point is possible, interpolating with the scaling form, which needs to be inverted again,
\begingroup
\allowdisplaybreaks
\begin{align}
    a\Tc(am,\Nf)
    &= aT_\tric(\Nf) + \coeff{E}{1}(\Nf) (am)^{2/5} + \coeff{E}{2}(\Nf)(am)^{4/5} + \ldots\;, \label{eq:T_m}\\[1ex]
    \Big(a\mc(\NTau,\Nf)\Big)^{2/5}
    &= \coeff{F}{1}(\Nf)\big(aT-aT_\tric(\Nf)\big) 
    + \coeff{F}{2}(\Nf)\big(aT-aT_\tric(\Nf)\big)^{2} + \ldots \;.\label{eq:m_T}
\end{align}
\endgroup
This gives the central line in \cref{fig:nf5-2nd}. Fits corresponding to the inverted \cref{eq:T_m} with $ \coeff{E}{1}=0$
or $\coeff{E}{2}=0$ bound this intercept and give a measure of uncertainty.
%A good fit to only the next-to-leading term of the scaling form is also possible, and describes the data well, \cref{fig:nf5-2nd}.
Our current data for $\Nf=5$ are thus consistent with a tricritical point at $\NTau^\mathrm{tric}(\Nf=5)\approx 12.5$,
and the same holds for $\Nf=6,7$ with slightly larger $\NTau^\mathrm{tric}$, cf.~\cref{fig:m-nt}.
This observation is fully compatible with the one made in \cref{fig:m-nf_all}: in the lattice chiral
limit $am=0$ there is a monotonically rising tricritical line $\Nf^\mathrm{tric}(\NTau)$. 

This has profound consequences for the approach to the continuum chiral limit. Avoiding lattice artefacts requires to 
take the continuum limit before the chiral limit. Since the continuum limit is represented by the origin in \cref{fig:m-nt},
the existence of a tricritical point $\NTau^\mathrm{tric}(\Nf)$ implies that the continuum chiral limit is inevitably approached
from the crossover region, and hence can only represent a second-order transition. In \cref{fig:m-nt} this appears to be 
the case for all $\Nf\in[2,7]$, which would then all feature second-order transitions in their respective continuum chiral limits.  

\section{Wilson fermions}

In view of our surprising findings from the last sections, it is particularly important to do a similar analysis in
a different discretisation scheme. To this end, we re-analyse already published data \cite{Jin:2014hea,Jin:2017jjp,Kuramashi:2020meg}
for the critical pseudo-scalar
mass delimiting the first-order transition for $\Nf=3$ $O(a)$-improved Wilson fermions with $\NTau\in[4,12]$. 
This is shown in \cref{fig:wilson}, where we have employed
$am_{PS}^2\propto am$ in order for the vertical axis to represent the scaling field, i.e.~the (additively renormalised) quark mass.

The lines in the figure represent leading-order scaling fits to $\NTau\in[8,12]$ and next-to-leading order scaling fits to 
$\NTau\in[6,12]$ as well as to $\NTau\in[4,8]$. 
An excellent description of the data is achieved in all three cases with $\chi^2_\mathrm{dof}$
near 1 and only small variation of the intercept, which represents a $\NTau^\mathrm{tric}(\Nf=3)$ for this 
non-perturbatively improved Wilson discretisation. This confirms the viability of our staggered analysis based on $\NTau\in[4,8]$,
and it leads to the same conclusion: the first-order chiral phase transitions observed  for $\Nf=3$ $O(a)$-improved Wilson 
fermions are not connected to the the continuum chiral limit, which therefore must represent a second-order transition.

\section{Conclusions}

\begin{figure}[t]
    \centering
    \begin{minipage}{0.5\textwidth}
       \centering
       \includegraphics[width=\linewidth]{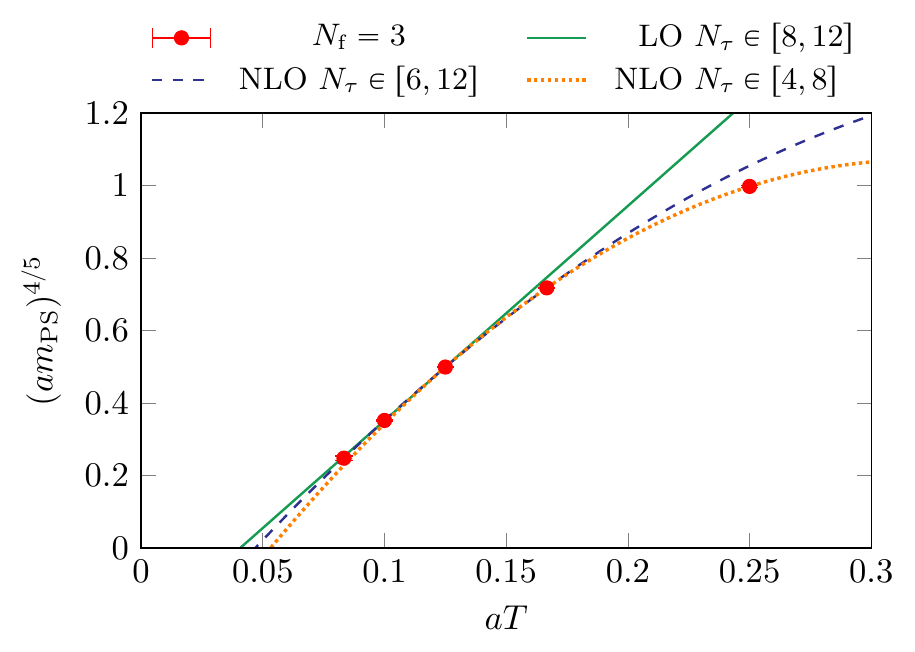}
       \captionof{figure}{
                Chiral critical line for $\Nf=3$ $O(a)$-improved Wilson fermions, with various fits to tricritical scaling.
                The data are taken from \cite{Kuramashi:2020meg}, the figure from \cite{Cuteri:2021ikv}.
            }\label{fig:wilson}  
     \end{minipage}%
     \hfill
     \begin{minipage}{0.45\textwidth}
       \centering
       \includegraphics[width=0.9\linewidth]{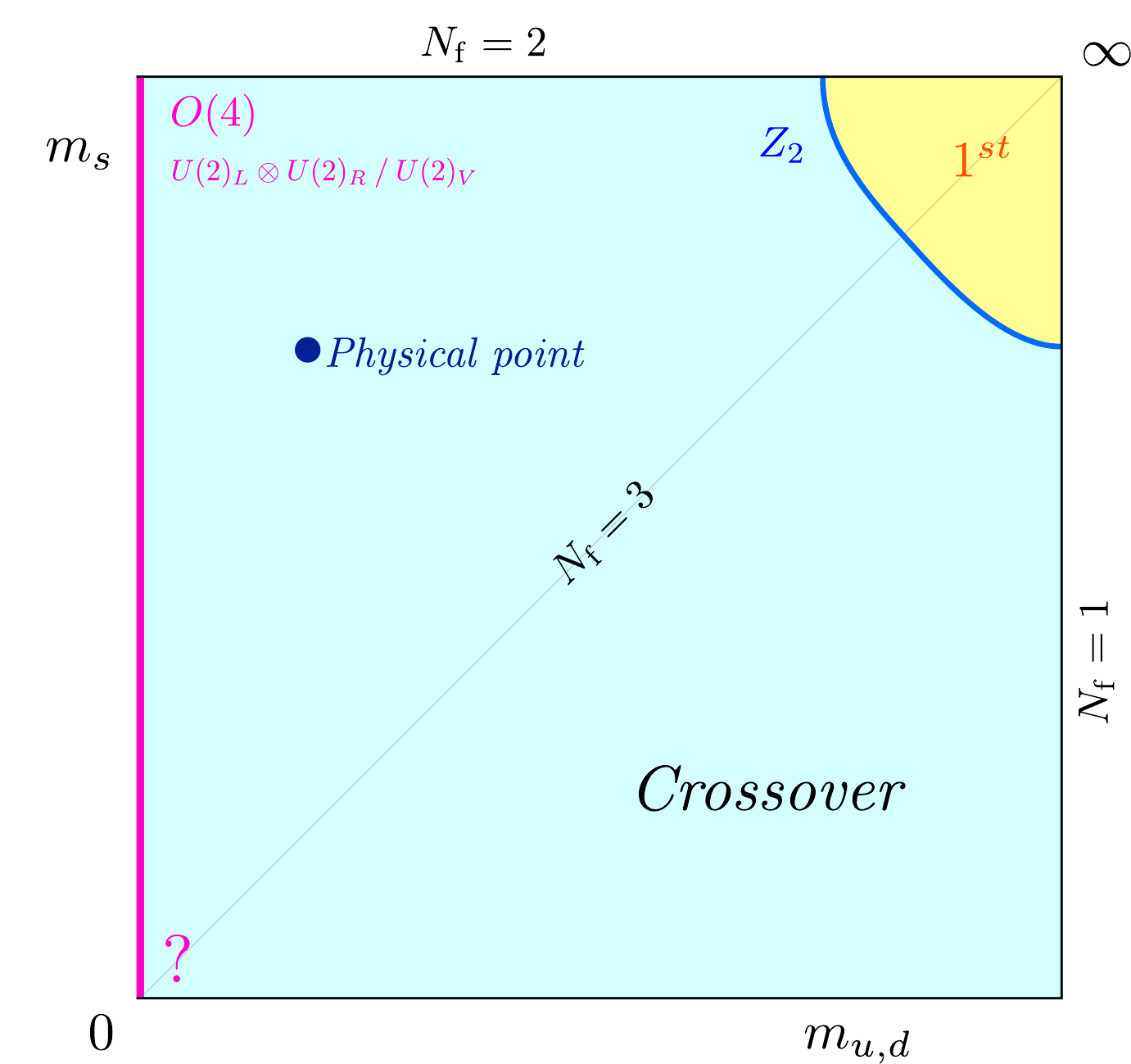}
       \captionof{figure}{
                Continuum Columbia plot suggested by our results. The chiral transition is of second order for 
                all values of $m_s$, its universality class is not determined here.  From \cite{Cuteri:2021ikv}.          
            }\label{fig:columbia-final}  
     \end{minipage}
\end{figure}
In summary, we have exploited the fact that a change of the massless chiral phase transition from first to second order as
a function of either $\Nf$ or lattice spacing must pass through a tricritical point. Its location can be determined by tricritical
scaling of the $Z_2$-critical line, which separates the first-order phase transitions from crossover behaviour and extrapolates
to a tricritical point in the chiral limit. A comprehensive analysis of this boundary for 
unimproved staggered quarks on lattices with $\NTau\in\{4,6,8\}$ is consistent with all $\Nf\in[2,7]$ displaying such a tricritical point.
This implies that the first-order region is not connected to the continuum limit which, therefore, must correspond to a second-order
transition. The same result is found for $\Nf=3$ $O(a)$-improved Wilson fermions. These conclusions can only be avoided
if future results for the chiral critical line on larger $\NTau$ break off the tricritical scaling curves. 
However, based on the currently available data one would have to conclude that the Columbia plot looks as in \cref{fig:columbia-final}.
All results reported here have been published in \cite{Cuteri:2021ikv}.
Finally, we note that our analysis can be applied to any discretisation with explicit first-order transitions.
This should allow to fully settle the question of the order of the massless chiral phase transition in the near future.

\acknowledgments
The authors acknowledge support by the Deutsche Forschungsgemeinschaft (DFG) through the 
grant CRC-TR 211 ``Strong-interaction matter under extreme conditions''. F.C.~and O.P.~in addition acknowledge support
 by the State of Hesse within the Research Cluster ELEMENTS (Project ID 500/10.006).

\bibliographystyle{JHEP}
\bibliography{bibliography}

\end{document}